\pdfoutput=1
\documentclass{caosp305}

\usepackage{graphicx}
\usepackage{upgreek}
\usepackage{natbib}
\articleNo{123}
\pubyear{2005}
\volume{35}
\volnumber{3}
\firstpage{1}
\received{May 1, 2005}
\accepted{August 28, 2005}

\def\BibTeX{{\rm B\kern-.05em{\sc i\kern-.025em b}\kern-.08em
             T\kern-.1667em\lower.7ex\hbox{E}\kern-.125emX}}

\begin{document}

%
\hauthor{A.~J. Martin}

\title{The evolution of magnetic fields from the main-sequence to very late stages}


%
\author{A.~J. Martin}

%
\institute{LESIA, Observatoire de Paris, PSL Research University, CNRS, Sorbonne Universit\'{e}s, UPMC Univ. Paris 06, 
Univ. Paris Diderot, Sorbonne Paris Cit\'{e}, 5 place Jules Janssen, F-92195 Meudon, France\\ \email{alexander.martin@obspm.fr}  }

\date{March 8, 2003}
\maketitle
\begin{abstract}
Magnetic fields have been detected in most if not all types of stars across the Hertzsprung-Russell diagram. Where present, these fields have the potential to significantly impact the evolution of their host stars. Furthermore, they themselves are affected by the various structural changes which occur in a star during its life. For example, the significant radius expansion during the post-main-sequence phase, due to flux conservation, may lead to a decrease in surface magnetic field strength, to a point where the magnetic field may no longer be detectable. As a result, it is a challenge to link the magnetic fields observed in main sequence (MS) stars with those observed in very late stage stars and even to those in post-MS stars. In this review, I present what we know, from observations, about magnetic stars at various stages of stellar evolution.\\
\end{abstract}
\section{Introduction}
The magnetic fields present in stars are shown to affect their host star 
in a variety of ways, both at the stellar surface and in the interior. For example, they modify  
the mixing and diffusion of chemical elements \citep{Stift2016} and the surface wind \citep{udDoula2002,udDoula2008,udDoula2009}. Thus, a good understanding of the structure of the 
magnetic field is important for the study of the evolution of magnetic stars.
However, the magnetic fields present in stars are themselves affected by the changing physical properties of 
the star as it undergoes the sometimes extreme structural changes during its life.
In this review, I explore what we infer from observations about the magnetic field evolution for stars at different 
stages of their lives, from the MS to the very late stages.
\section{Dynamo and fossil fields}
In general, we infer two main mechanisms for the presence of magnetic fields in stars. Those of ``dynamo origin'' 
\citep[e.g.,][]{Charbonneau2014} are magnetic fields generated as the result of a dynamo process in the inner layers of 
a star. Those of ``fossil field origin'' \citep[e.g.,][]{Braithwaite2004} are magnetic fields inherited and possibly amplified in the earlier stages of a star's life. Since the mechanism by which these fields form is different, different processes can affect their evolution.
\subsection{Ohmic decay}

In the absence of any additional degenerative or regenerative force a 
magnetic field will decay over a time-scale called the Ohmic decay time, $t_{\rm ohm}$.
This is given by
\begin{equation}
t_{\rm ohm} = \frac{4\pi\sigma L^2}{c^2},
\end{equation}
where L is the magnetic field length scale, c is the speed of light and 
$\sigma$ is the electrical conductivity. In the context of this review, Ohmic decay is 
only relevant for fossil fields, since dynamo fields are constantly regenerated.
The Ohmic decay time gives a baseline for the time over which we could expect a field 
to decay. This means we can investigate whether the field is decaying faster and so by 
a different means.
\subsection{Magnetic flux conservation}

If there is no change in magnetic field flux then, as a star evolves and its 
radius expands or contracts, by the laws of flux conservation the surface 
magnetic field strength is expected to decrease or increase respectively. 
The current dipole magnetic field strength ($B_{\rm d, current}$) after a change 
in radius is calculated as
\begin{equation}
B_{\rm d, current} = B_{\rm d, previous}\left(\frac{R_{\rm previous}}{R_{\rm current}}\right)^2,
\end{equation}
where $B_{\rm d, previous}$ is the previous dipole magnetic field strength, $R_{\rm previous}$ is the previous radius of the star and 
$R_{\rm current}$ is the current radius of the star.

\section{Magnetic field evolution during the main sequence}
During the evolution of stars along the MS, two key properties change. 
First the rotation period of the star increases as a result of angular 
momentum loss and second in all but the smallest of stars the 
radius increases. In stars of approximately one solar mass, this is rather modest 
and between the zero-age main-sequence (ZAMS) and terminal-age main-sequence (TAMS) 
we would expect a radius increase of a factor of $\sim 1.4$ based on the evolutionary models of \citet{Ekstrom2012}. 
In contrast, for stars with M $ = 50\,{\rm M}_\odot$ we would expect a radius increase of a factor of $\sim 6$ based on the evolutionary models of \citet{Ekstrom2012}.  In addition, we expect that the time that a star spends on the MS decreases 
with increasing ZAMS mass.

\subsection{Low-mass stars}

Stars with a masses below $\sim 1.5$M$_\odot$ have been shown to host dynamo magnetic fields \citep[e.g.,][]{Donati2011}. 
\citet{Vidotto2014} investigated the prediction of \citet{Skumanich1972} that a dynamo magnetic field will 
decay as the inverse square of stellar age. Their sample has a mass range of 0.1--2.0\,M$_\odot$ and an 
age range of 1\,Myr--$\sim$10\,Gyr. They find a power law fit $\langle|B_V|\rangle\propto t^{-0.655\pm0.045}$ which supports 
the prediction of \citet{Skumanich1972}. In addition, they find a similar relationship between the unsigned surface flux and stellar 
age: $\Phi_V \propto t^{-0.622\pm0.042}$. 
\begin{figure}
\includegraphics[width=\textwidth]{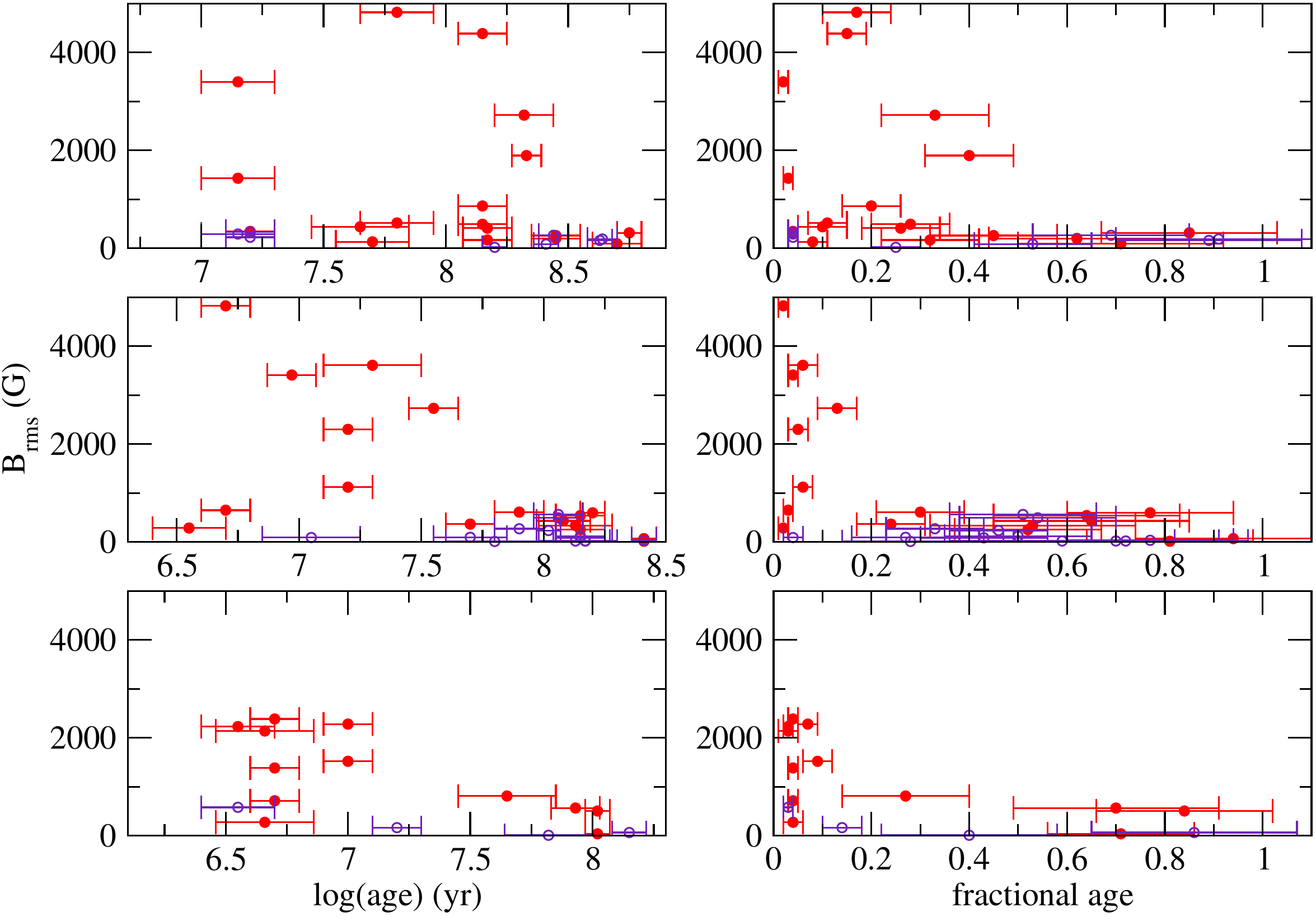}
\caption{$B_{\rm rms}$ as a function of logarithmic stellar age (left) and as a function of 
fractional MS age (right) for the sample of stars given in \citet{Landstreet2008}. Filled symbols = stars with definite field detections, open symbols = probable Ap stars for which a magnetic field has not yet been detected. From top to bottom the panels indicate three mass bins, 2--3M$_\odot$, 3--4M$_\odot$, and 4--5M$_\odot$. 
Credit: Landstreet, et al., A\&A, 481, 465, 2008, reproduced with permission \textcopyright\ ESO.}
\label{fig:Landstreet1}
\end{figure}
\subsection{Intermediate-mass stars}
Stars with masses larger than  $\sim$1.5\,M$_\odot$ host fossil magnetic fields. The incidence rate of magnetic fields in Ap/Bp stars is $\sim$10\% and they host large-scale magnetic fields with strengths ranging 
from 300\,G to 30\,kG \citep{Power2008,Auriere2007}. 
\citet{Landstreet2008} studied the magnetic fields of 23 late B- and early A-type stars and their results are shown in  Fig.~\ref{fig:Landstreet1} and their Fig.~5. They find that the magnetic field of stars with masses between 2 and 3\,M$_{\odot}$ declines after about 2.5 $\times$ 10$^8$\,yr, for 
3-4\,M$_\odot$  the field declines after about     4$\times$ 10$^7$\,yr and for 
4-5\,M$_\odot$  the field declines after about  1.5$\times$ 10$^7$\,yr as shown in Fig.~\ref{fig:Landstreet1}. 
In addition, as shown by their Fig.~5, \citet{Landstreet2008} find that magnetic flux decays with stellar age for all of their sample. 
They conclude 
that either magnetic flux declines in all Ap/Bp stars during the MS lifetime or that magnetic flux in the more 
strongly magnetic stars found at young fractional ages is somehow reduced.
\subsection{Massive stars}
\begin{figure}
\includegraphics[width=.5\textwidth]{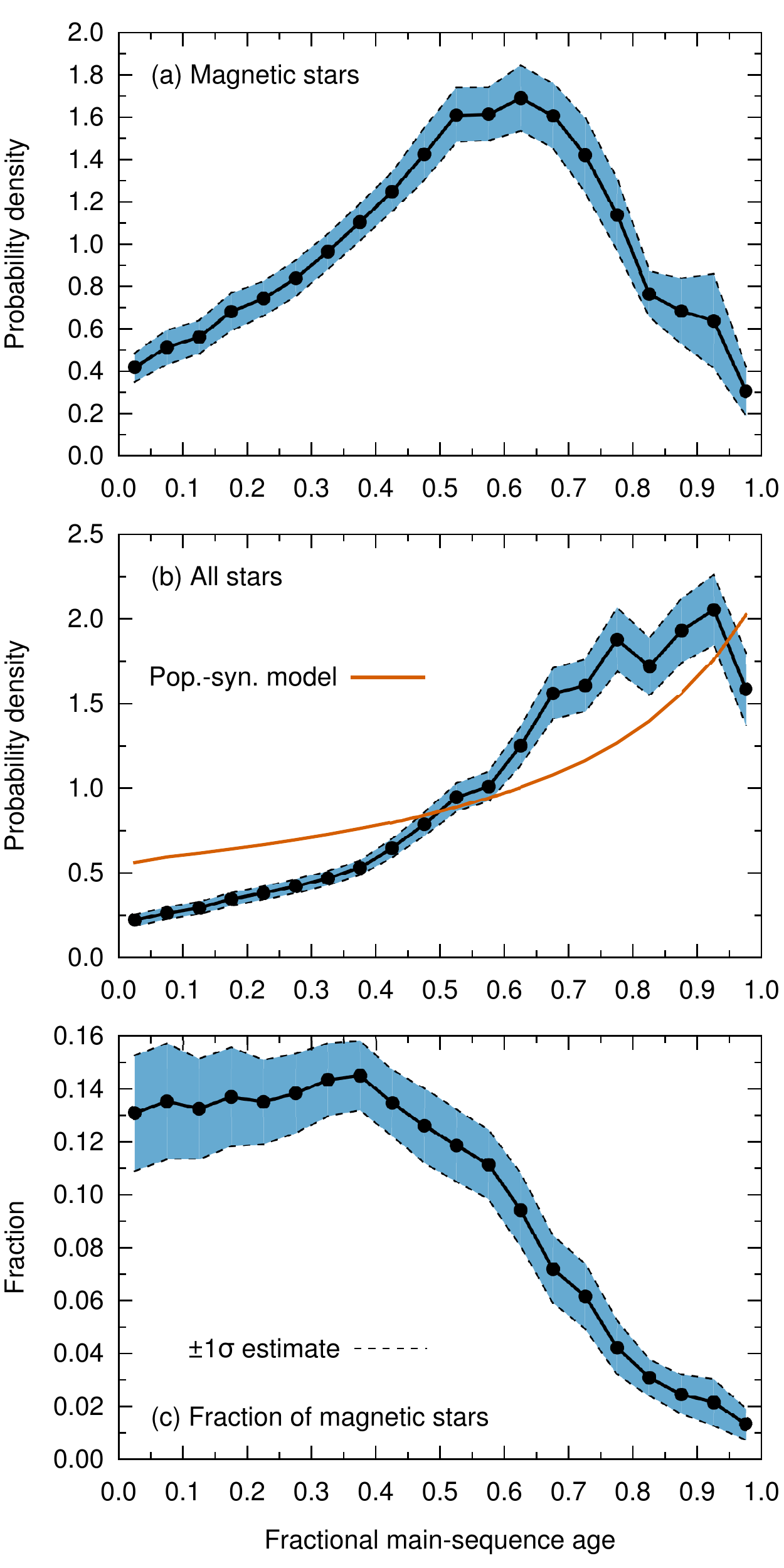}\includegraphics[width=.5\textwidth]{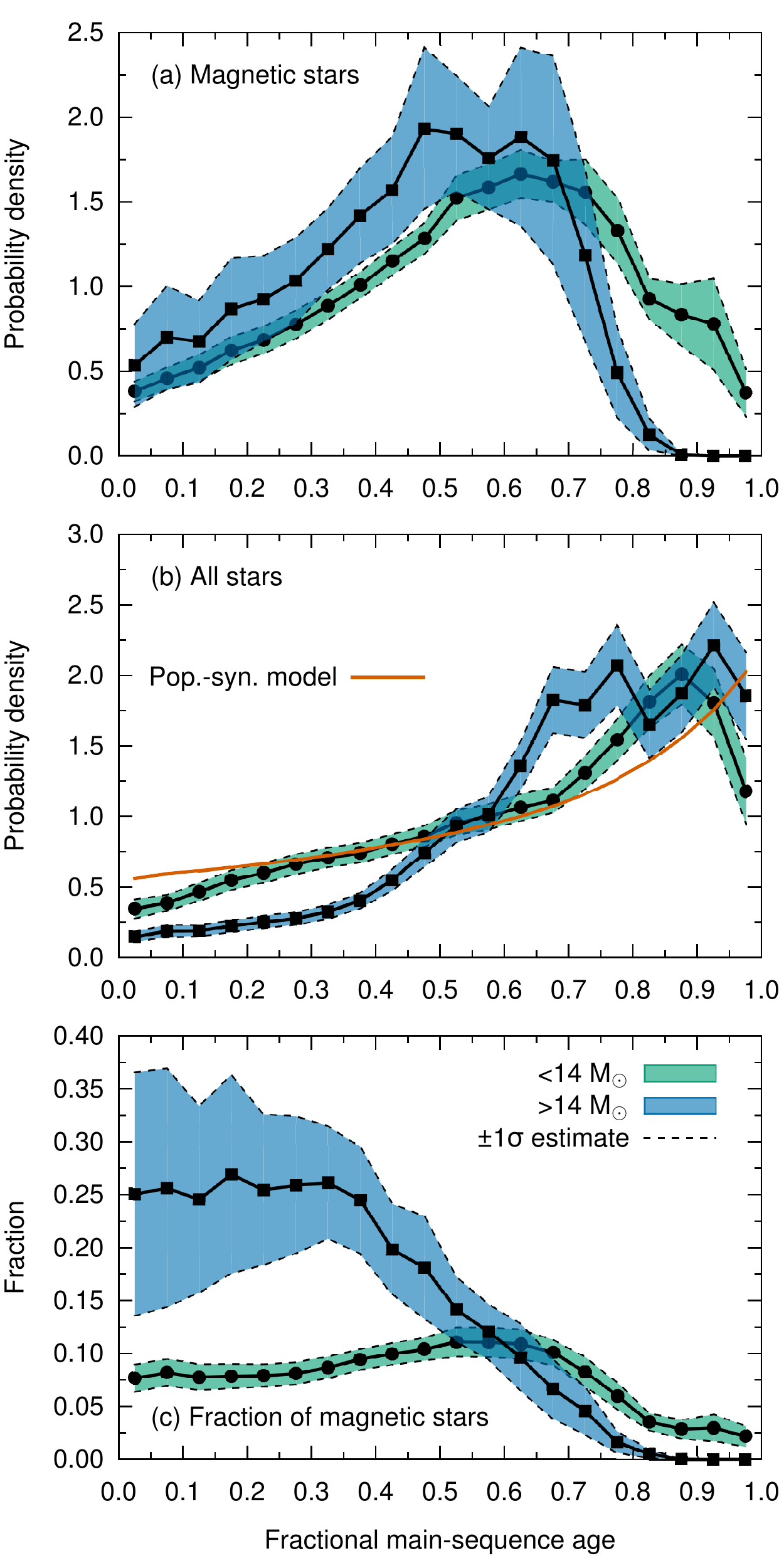}
\caption{
``Fractional MS $\uptau$-distributions of (a) the magnetic stars and (b)
all stars in the sample from \citet{Fossati2016}. Panel (c) shows the fraction of magnetic stars
as a function of fractional MS age ($\uptau$), normalised such that the incidence
of magnetic stars is 7\% (Wade et al. 2014; Fossati et al. 2015b). The
shaded regions indicate bootstrapped 1$\upsigma$ estimates to give an indication
of the statistical significance of the variability in the $\uptau$-distributions. The
solid line in the middle panel shows the $\uptau$-distribution of a synthetic
population of a magnitude-limited sample of massive stars (see the text of \citet{Fossati2016}).''
Figure and caption (adapted) credit: Fossati, et al., A\&A, 592, A84, 2016, reproduced with permission \textcopyright\ ESO.}
\label{fossati}
\end{figure}
Like the intermediate-mass stars, there is convincing evidence that $\sim$\,10\,\% 
of massive stars host a fossil magnetic field \citep{Grunhut2015,Fossati2015,Wade2016,Grunhut2017}. The study by 
\citet{Fossati2016} investigated the incidence of magnetic field as a function of fractional MS age ($\uptau$). Their results are shown in 
Fig.~\ref{fossati}. They find that the incident rate of magnetic fields in massive stars sharply decreases above $\uptau \sim 0.6$, with this 
effect being significantly more pronounced for stars above 14\,M$_\odot$. \citet{Fossati2016} investigated the possible causes of this, and conclude that 
binary rejuvenation and suppression of core convection cannot explain the observed distributions. In addition, while flux 
conservation is likely a contributing factor, if it were the sole explanation for this, at least a third of the stars which have a detectable 
magnetic field on the ZAMS, would still be detectable as magnetic on the TAMS. As a result, they conclude that there is likely 
a mass dependant field decay mechanism operating.
\section{Post-main sequence stars}
The post-main sequence of most stars is characterised by significant inflation of the stellar radius and the end of H-burning. Therefore, for stars with 
fossil magnetic fields, it is predicted that as a result of flux conservation the surface magnetic field strength will be significantly reduced. 
\subsection{Post-MS FGK-type stars}
The study by \citet{Grunhut2010} determined that a third of their sample of $\sim$30 massive late-type stars host dynamo magnetic fields. To explain such a high incidence rate in comparision to other stars, they conclude that a mechanism may exist to excite  magnetic fields in cool supergiants. The work by \citet{Auriere2015} investigated the magnetic fields of active G-K giants. They found that most of the stars in their sample were either in the first dredge-up phase or core helium-burning phase. As a result they conclude that this could be evidence for a  magnetic strip of the most active stars.
\subsection{Post-MS OBA stars and the LIFE project}
\begin{figure}
\includegraphics[width=0.48\textwidth]{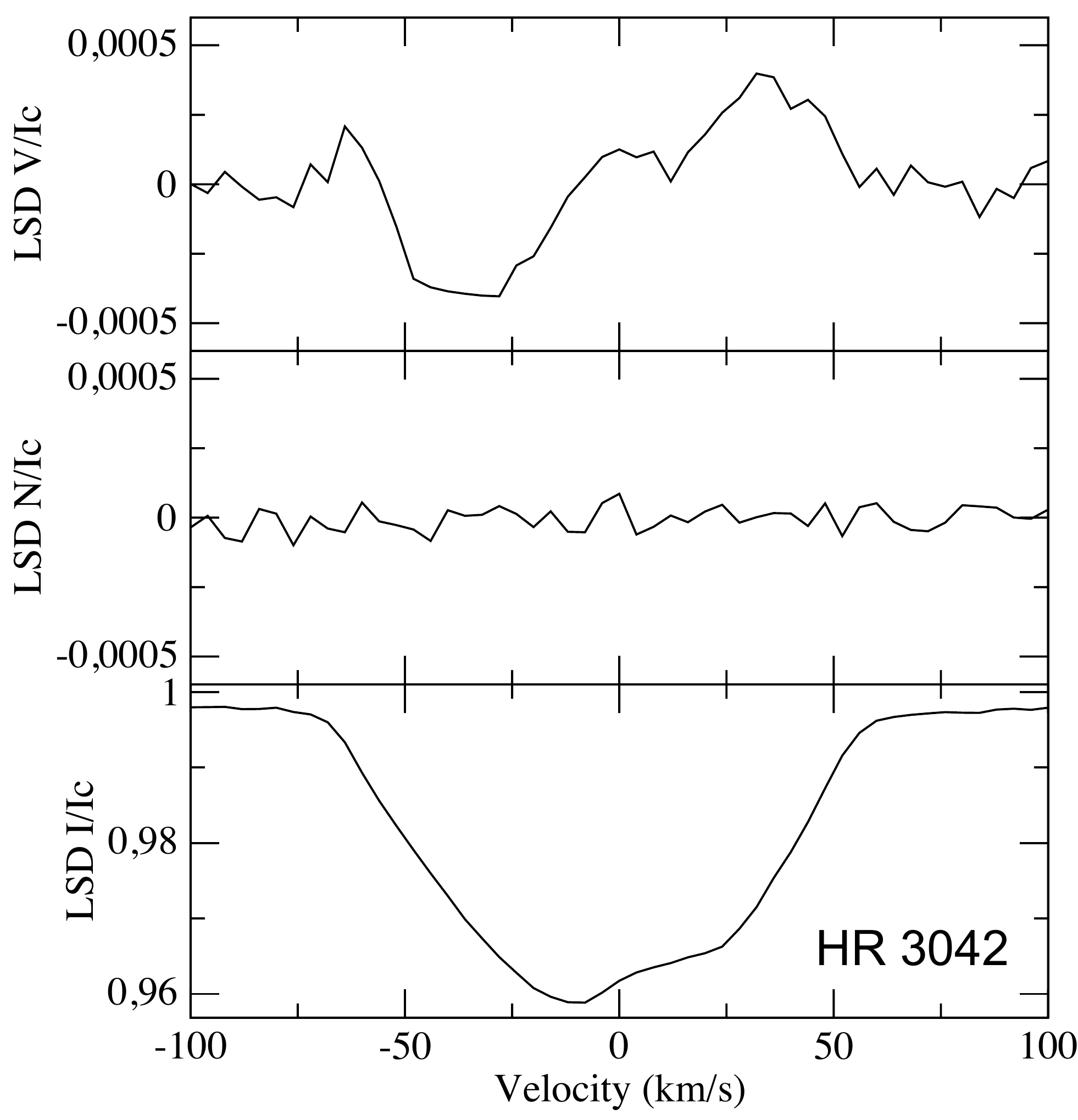}\hfill\includegraphics[width=0.48\textwidth]{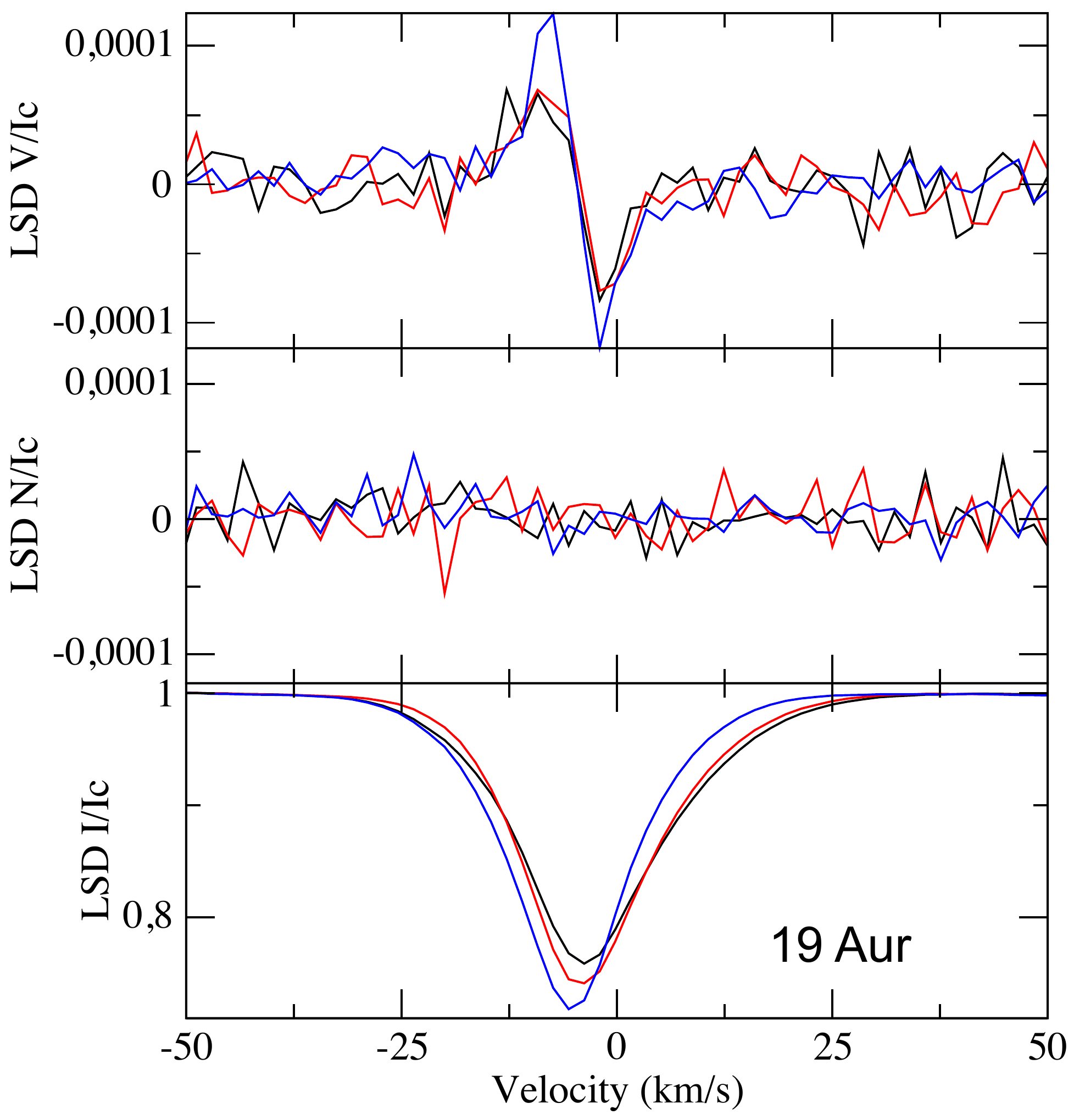}
\caption{LSD Stokes $V$ (top panels), Null (middle panels) and Stokes $I$ (bottom panels) profiles of the B8III star HR\,3042 with $B_\ell=-230\pm10\,$G (left) and the A5Ib-II star 19\,Aur with $B_\ell=1.0\pm0.2\,$G (right; 3 observations superposed)
 obtained by the ESPaDOnS instrument of the CFHT in semester 2016B, showing that these 2 stars are magnetic (Martin et al., submitted).}
\label{LIFE}
\end{figure}
Until recently there were no known, clearly evolved, magnetic hot stars. However, as part of the 
BRIght Target Explorer spectropolarimetric survey  
\citep[BRITEpol;][]{Neiner2016}, \cite{Neiner2017} identified two magnetic A7 supergiants: $\upiota$\,Car and HR\,3890.
To systematically study the magnetic fields of hot giants and supergiants, the Large Impact of magnetic Fields on the Evolution of hot stars (LIFE) project (Martin et al., submitted) was started. These stars provide an important link between magnetic stars on the MS and the late stages of a star's life. So far, 15 stars have been observed as part of this project. Out of these, two have been found to be magnetic: HR\,3042  and 19\,Aur (see Fig.~\ref{LIFE}). The current strength of the magnetic fields detected in these stars is compatible with what we would expect if conservation of magnetic flux was the main driver for magnetic field evolution, however, 
it does not exclude the possibility of other magnetic field decay processes affecting the field evolution.
\section{Late stages of stellar evolution}
The late stages of stellar evolution are characterised by arguably the most extreme changes to occur in 
the stellar structure since the birth of the star. As a result, it is challenging to link the magnetic field observed in these stars with those 
observed in MS stars.
\subsection{White dwarfs}
White dwarfs (WDs) are the final evolutionary stage of most stars: those stars with 
masses less than $\sim$8\,M$_\odot$. They are found to have a broad range of magnetic field strengths from $\sim$10$^3$\,G to 10$^9$\,G \citep[e.g.,][]{Ferrario2015}. The lower limit of $\sim$10$^3$\,G may, however, be the result of instrument limitations rather than a true lower limit, in which case, the range in field strengths could be even greater. The mechanism responsible for the presence of magnetic fields in WDs remains unclear, however, possible explanations include: they are fossil fields preserved from the MS; they are fossils of the strong convective core fields of red giants, revealed as the outer layers of the star are stripped off; or they are fossils of dynamo fields generated in common envelope evolution of close binaries. Furthermore, It is interesting to note that currently the incidence rate of magnetic fields in WDs 
appears to be $\sim$10\% \citep{Ferrario2015}, which is consistent with the incidence rate seen for the hot stars  
 mentioned earlier in this review. A comprehensive review by \citet{Ferrario2015} shows the 
results of the magnetic field measurement of a large sample of WDs in their Fig.~3. This figure shows that there 
is no evidence for magnetic field evolution with age, which is consistent with the long ohmic decay timescale (2--6 $\times$ 10$^{11}$ yr) for WDs.
\subsection{Neutron Stars}
The final evolutionary stage of more massive stars from $\sim 8$\,M$_\odot$ to $\sim 25-30$\,M$_\odot$ is the neutron star, formed as 
the result of a supernova. Neutron stars are found to have magnetic field strengths of 
$\sim10^8-10^{14}$\,G \citep[e.g.,][]{Harding2006}, where the strongest magnetic fields of  $\sim10^{14}$\,G are above the quantum critical field strength. The study by \citet{Harding2006} shows that the magnetic fields of these stars decrease in strength with stellar age in their Fig.~1. However, connecting the magnetic fields observed in neutron stars with those observed on the MS is a 
challenge, because of the extreme physics acting during their birth. The continued analysis of magnetic fields of stars at various evolutionary stages is therefore necessary to connect the fields we observe on the MS with those at the very late stages of stellar life.

\acknowledgements
I am very grateful to John Landstreet, Stefano Bagnulo, Coralie Neiner and Luca Fossati for their help collating the information necessary for 
this review.

\bibliography{Alex_Martin}

\end{document}